\documentclass[a4paper,fleqn,usenatbib]{mnras}

\usepackage{natbib}
\usepackage{soul}

\usepackage{newtxtext,newtxmath}

\usepackage[T1]{fontenc}
\usepackage{ae,aecompl}

\usepackage{todonotes}
\usepackage{outlines}
\usepackage{color}

\usepackage{graphicx}	
\usepackage{amsmath}	
\usepackage{bm}	        

\AtBeginShipout{%
  \ifnum\value{page}>1 %
    \typeout{* Additional boxing of page `\thepage'}%
    \setbox\AtBeginShipoutBox=\hbox{\copy\AtBeginShipoutBox}%
  \fi
}
\def\xihm{\xi_{\mathrm{hm}}}
\def\ximm{\xi_{\mathrm{mm}}}
\def\mrm{\mathrm}

\newcommand{\lcdm}{\Lambda\mathrm{CDM}}
\newcommand{\avg}[1]{\left\langle #1 \right\rangle}

\newcommand{\hmsol}{h^{-1}M_{\odot}}

\newcommand{\msbcg}{M_*^{\mathrm{BCG}}}
\newcommand{\mh}{M_h}
\newcommand{\mpc}{\mathrm{Mpc}}
\newcommand{\hmpc}{h^{-1}{\mathrm{Mpc}}}

\newcommand{\hkpc}{h^{-1}{\mathrm{kpc}}}

\newcommand{\hhmsol}{h^{-2}M_{\odot}}
\newcommand{\ds}{\Delta\Sigma}

\newcommand{\kms}{{\mathrm{km}}\,s^{-1}}

\newcommand\redmapper{redMaPPer}

\newcommand{\rom}[1]{\uppercase\expandafter{\romannumeral #1\relax}}

\title[Scatter in the Cluster SHMR]{Does Concentration Drive the Scatter in the Stellar-to-Halo Mass Relation of Galaxy Clusters?}
\author[Zu 2020]{Ying  Zu$^{1,2}$\thanks{E-mail: yingzu@sjtu.edu.cn},
Huanyuan Shan$^{3}$, Jun Zhang$^{1,2}$,
Sukhdeep Singh$^{4}$, Zhiwei Shao$^{1}$,
\newauthor
Xiaokai Chen$^{1}$, Ji Yao$^{1}$,
Jesse B. Golden-Marx$^{1}$,
Weiguang Cui$^{5}$,
Eric Jullo$^{6}$,
\newauthor
Jean-Paul Kneib$^{6, 7}$,
Pengjie Zhang$^{1,2}$,
Xiaohu Yang$^{1,2}$ \\ \\
$^{1}$Department of Astronomy, School of Physics and Astronomy, Shanghai Jiao Tong
University, Shanghai 200240, China\\
$^{2}$Shanghai Key Laboratory for Particle Physics and Cosmology, Shanghai Jiao Tong
University, Shanghai 200240, China\\
$^{3}$Key Laboratory for Research in Galaxies and Cosmology, Shanghai Astronomical Observatory, Shanghai 200030, China \\
$^{4}$McWilliams Center for Cosmology, Department of Physics, Carnegie Mellon University,
5000 Forbes Avenue, Pittsburgh, PA 15213, USA\\
$^{5}$Institute for Astronomy, University of Edinburgh, Royal Observatory, Edinburgh EH9
3HJ, United Kingdom\\
$^{6}$Aix-Marseille Univ, CNRS, CNES, LAM, Marseille, France\\
$^{7}$Institute of Physics, Laboratory of Astrophysics, Ecole Polytechnique
F\'{e}d\'{e}rale de Lausanne (EPFL), Observatoire de Sauverny, 1290 Versoix, Switzerland
}

\date{Accepted XXX. Received YYY; in original form ZZZ}

\pubyear{2020}

\begin{document}

\label{firstpage}
\pagerange{\pageref{firstpage}--\pageref{lastpage}}
\maketitle

\begin{abstract} Concentration is one of the key dark matter halo properties that could
    drive the scatter in the stellar-to-halo mass relation of massive
    clusters. We derive robust photometric stellar masses for a sample of
    brightest central galaxies~(BCGs) in SDSS redMaPPer clusters at
    $0.17{<}z{<}0.3$, and split the clusters into two equal-halo mass subsamples
    by their BCG stellar mass $\msbcg$. The weak lensing profiles $\ds$ of the
    two cluster subsamples exhibit different slopes on scales below $1\hmpc$.
    To interpret such discrepancy, we perform a comprehensive Bayesian modelling
    of the two $\ds$ profiles by including different levels of miscentring
    effects between the two subsamples as informed by X-ray observations. We
    find that the two subsamples have the same average halo mass of
    $1.74\times10^{14}\hmsol$, but the concentration of the low-$\msbcg$
    clusters is $5.87_{-0.60}^{+0.77}$, ${\sim}1.5\sigma$ smaller than that of their
    high-$\msbcg$ counterparts~($6.95_{-0.66}^{+0.78}$). Furthermore, both
    cluster weak lensing and cluster-galaxy cross-correlations indicate that the
    large-scale bias of the low-$\msbcg$, low-concentration clusters are ${\sim}10\%$
    higher than that of the high-$\msbcg$, high-concentration systems, hence
    possible evidence of the cluster assembly bias effect. Our results reveal a
    remarkable physical connection between the stellar mass within $20{-}30\hkpc$, the
    dark matter mass within ${\sim}200\hkpc$, and the cosmic overdensity on
    scales above $10\hmpc$, enabling a key observational test of theories of
    co-evolution between massive clusters and their central galaxies.
\end{abstract}




\vspace{1in}
\section{Introduction}
\label{sec:intro}

As the most dominant galaxies at the center of massive
clusters~\citep{Kravtsov2012, VonDerLinden2007}, the brightest central
galaxies~(BCGs\footnote{We use BCGs to refer to the brightest central or
brightest cluster galaxies interchangeably.}) have witnessed both the early-time
fast growth and the late-time slow accretion of cluster haloes predicted by
hierarchical structure formation in the $\lcdm$ Universe~\citep{Zhao2003,
Klypin2016}. Consequently, the BCGs have likely experienced an analogous
two-phase formation, with starbursts induced by gas-rich
mergers~\citep{Barnes1991, Mihos1996, Hopkins2013} and fast
accretion~\citep{Fabian1994, Collins2009, McDonald2012} at the onset of cluster
formation, followed by dry mergers and smooth accretions until the observed
epoch~\citep{Lin2004, Bezanson2009, vanDokkum2010, Zhang2016,
Huang2018, DeMaio2020}. The correlated growth of cluster haloes and BCGs is also
seen in numerical simulations of cluster formation~\citep{Dubinski1998,
Mostoghiu2019, Ragagnin2019, Rennehan2020}.  Therefore, the scatter in
the stellar-to-halo mass relation of BCGs inevitably carries the imprint of the
assembly history of cluster haloes, providing an important avenue to the
physical understanding of galaxy-halo connection at the very massive
end~\citep{Wechsler2018}. In this paper, we measure the weak gravitational
lensing signals of two equal-halo mass cluster subsamples split by BCG stellar
mass, and look for possible discrepancies in their halo concentration $c$, one
of the most fundamental halo properties that correlates strongly with halo
assembly history.

The mean concentration of haloes declines slowly with mass in the cluster
regime~\citep{Jing2000, WangJie2020}, with a hint of an upturn at the highest mass
end~\citep{Klypin2011, Prada2012}. Using the extended Press-Schechter
formalism~\citep{Bond1991, Bower1991, Lacey1993}, \citet{Navarro1997} was the
first to suggest that $c$ is linked to the formation time of haloes. Since then,
the physical connection between the $c-\mh$ relation and the average mass
assembly history~(MAH) of haloes has been the subject of extensive
study~\citep{Bullock2001, Wechsler2002, Zhao2009, Ludlow2013, Diemer2015, Ishiyama2020}. For example,
\citet{Salvador-Sole1998} found that the scale radius of haloes is essentially
proportional to their virial radius at the time of formation.  \citet{Zhao2003}
found that the MAH generally consists of an early phase of fast accretion when
$c$ stays roughly constant and a late phase of slow accretion when $c$ increases
with time, separated at a time when $c{\sim}4$ and the typical binding energy of
the halo is approximately equal to that of a singular isothermal sphere with the
same circular velocity.

However, haloes at fixed mass have significant scatter in their concentration
values~\citep{Jing2000}, and many studies have attempted to identify the key
parameter that causes individual haloes to deviate from the mean $c{-}\mh$
relation. Mergers strongly perturb individual halo formation histories from the
average MAH, but they mostly contribute to the variance of concentration at
fixed formation time and halo mass~\citep{Rey2019, Wang2020, Chen2020}. Applying
the secondary infall model~\citep{Gunn1972, Fillmore1984, Bertschinger1985} to
the initial Lagrangian region around cluster-size haloes, \citet{Ascasibar2007}
argued that the diversity in halo concentration arises from the scatter in the
primordial spherically average density profile rather than in the angular
momentum distribution of dark matter. Building on the insight from the peak
background split theory~\citep{Mo1996, Sheth1999}, \citet{Dalal2008} proposed
that the most important parameter is the curvature of the rare density peaks in
the initial Gaussian random field, so that highly curved peaks produce more
concentrated haloes. In a similar vein, \cite{Diemer2015} found that the
deviations can be explained by the residual dependence of $c$ on the local slope
of the matter power spectrum. Therefore, the scatter of concentration at fixed
mass is largely driven by the diversity of the local environments of the initial
Lagrangian peaks~\citep[including geometric environments; see][]{Hellwing2020}.

It is possible that the deviations of cluster haloes from the mean $c-\mh$
relation translate at least partly into the scatter in the stellar-to-halo mass
relation~(SHMR) of the BCGs.  Through starbursts induced by major mergers and
rapid accretion, the early growth of the BCGs at fixed halo mass should depend
critically on the strength of the potential well in the inner region of
clusters, which was already set before the scale radius~(defined as halo radius
over $c$) was stabilized~\citep{vandenBosch2017}. Therefore, the initial scatter
in BCG stellar mass should be tightly correlated with the MAH of haloes, hence
concentration. In the presence of strong Active Galactic Nuclei~(AGN)
feedbacks~\citep{Martizzi2012, Cui2018}, the BCG {\it in-situ} stellar growth
due to star formation would be quenched while the halo mass (and the {\it
ex-situ} component of BCG stellar mass) continues to grow, weakening the
correlation between BCG stellar mass and halo concentration.  However, AGN
feedback is linked to the growth history of the supermassive black hole~(SMBH),
which could also be correlated with concentration at fixed halo mass, in a
similar spirit to the $M_{\mathrm{BH}}$-$\sigma$ relation~\citep{Gebhardt2000,
Ferrarese2000, Gultekin2009}.  Indeed, hydro-simulations and semi-analytic
models that turn on AGN feedback in galaxy formation generally predict that the
scatter in SHMR is tied to halo formation time or concentration, albeit with
large uncertainties in the cluster mass range~\citep{Wang2013, Matthee2017,
Tojeiro2017, Artale2018, Zehavi2019, Bose2019}.  Conversely, the redistribution
of binding energy between dark matter and baryons during BCG formation may
slightly increase the halo concentration~\citep{Rudd2008}.

The observed scatter in the SHMR of clusters is generally believed to be
${\leq}0.2$ dex.  For example, \citet{Kravtsov2018} measured the scatter to be
${\sim}0.2$ dex from 21 X-ray clusters with individual halo mass estimated from
X-ray observations~\citep{Kravtsov2006}.  \citet{Zu2015} detected a weak
decreasing trend of scatter from $0.22$ dex in $L_*$-galaxies to $0.18$ dex in
the cluster regime. Finding the physical driver for such a small scatter
requires a large volume-limited sample of clusters with accurate measurements of
halo mass and concentration, which can then be divided into subsamples of
different average BCG stellar masses at fixed halo mass~\citep[but see][for
unbinned statistical methods]{GoldenMarx2018, GoldenMarx2019}. This has recently
become possible with the advent of large optical cluster samples detected from
all-sky imaging surveys~\citep{Kochanek2003, Koester2007, Szabo2011, Wen2012,
Rykoff2014, Oguri2018}. In particular, we will employ the volume-limited
\redmapper{} cluster~(\texttt{v6.3}) catalogue derived from the SDSS DR8
photometry~\citep{Rykoff2014}, and measure the average halo mass and
concentration from the stacked weak gravitational lensing profiles of clusters.

This paper is organized as follows. We describe the cluster catalogue, BCG
stellar mass estimates, and weak lensing measurements in~\S\ref{sec:data}. The
theoretical model of weak lensing signals and the Bayesian inference method are
described in~\S\ref{sec:bayes}. We present our model constraints on halo mass,
concentration, and bias in ~\S\ref{sec:results} and provide further measurement
of halo bias from cluster-galaxy cross-correlations in~\S\ref{sec:cross}. An
apparent conundrum on the SHMR at fixed satellite richness is discussed
in~\S\ref{sec:puzzle} before we conclude by summarising our results and look to
the future in ~\S\ref{sec:conc}.

Throughout this paper, we assume the {\it Planck} cosmology~\citep{Planck2018}.
All the length and mass units in this paper are scaled as if the Hubble constant
is $100\,\kms\mpc^{-1}$. In particular, all the separations are co-moving
distances in units of $\hmpc$, and the halo and stellar mass are in units of
$\hmsol$ and $\hhmsol$, respectively. We use $\lg x{=}\log_{10} x$ for the
base-$10$ logarithm and $\ln x{=}\log_{e} x$ for the natural logarithm.

\section{Data and Measurements}
\label{sec:data}

\subsection{Cluster Catalogue}
\label{subsec:cluster}

We employ the optical cluster catalogue derived from SDSS DR8~\citep{York2000,
Aihara2011} imaging using the red-sequence-based matched-filter photometric
cluster finding algorithm redMaPPer~\citep{Rykoff2014}. For each cluster, the
redMaPPer algorithm measures a richness $\lambda$ as its proxy for halo mass,
which corresponds roughly to the number of satellite galaxies brighter than
$0.2\,L_*$ within an aperture ${\sim}1\,\hmpc$~(with a weak dependence on
$\lambda$). At $\lambda{\geq}20$, the SDSS \redmapper{} cluster catalogue is
approximately volume-complete up to $z{\simeq}0.33$, with cluster photometric redshift
uncertainties as small as $\delta(z)=0.006/(1+z)$~\citep{Rykoff2014, Rozo2015}.
More important, the log-normal scatter of $\lambda$ at fixed halo mass is as
small as $0.2$-$0.25$ dex, providing an excellent halo mass proxy for our
analysis~\citep{Simet2017, Murata2018, Costanzi2019}.

To ensure that the cluster sample is volume-complete, we select the $4567$ BCGs in
the redMaPPer $\lambda{>}20$ clusters between
$z{=}0.17{-}0.30$~($\avg{z}{=}0.242$), within the same redshift range of the
BOSS LOWZ spectroscopic galaxy sample. Those BCGs are identified by the \redmapper{}
algorithm as the most likely central galaxies of the clusters.
Among the $4567$ BCGs, $957$ of them~(21
per cent) do not have spectroscopic coverage from SDSS, hence no
spectroscopic stellar mass measurements from, e.g., \citet{Chen2012}.
Unfortunately, those $957$ BCGs are preferentially systems with stellar mass~(as will be
estimated later with from broad-band photometry in ~\S~\ref{subsec:bcg})
below $10^{11}\hhmsol$ and above $3{\times}10^{11}\hhmsol$ --- they are the
either the lowest or the highest-$\msbcg$ galaxies at fixed
$\lambda$~\citep{Zu2020}. This spectroscopic incompleteness significantly
reduces our capability of resolving the full range of scatter in the cluster
SHMR. Additionally, the selection of the BOSS LOWZ galaxies relies on a complex
set of colour cuts~\citep{Reid2016}, which further complicates the selection
function of our otherwise volume-completed cluster sample. To circumvent the
spectroscopic stellar mass-incompleteness issue, we choose to re-measure the
stellar masses for all $4567$ BCGs using photometry. Thanks to the accurate
cluster photo-z estimates, we are able to derive the stellar masses with
reasonable accuracy from SDSS broad-band photometry in the next Section.

\subsection{Stellar Mass of the Brightest Central Galaxies}
\label{subsec:bcg}

\begin{figure}
\begin{center}
    \includegraphics[width=0.48\textwidth]{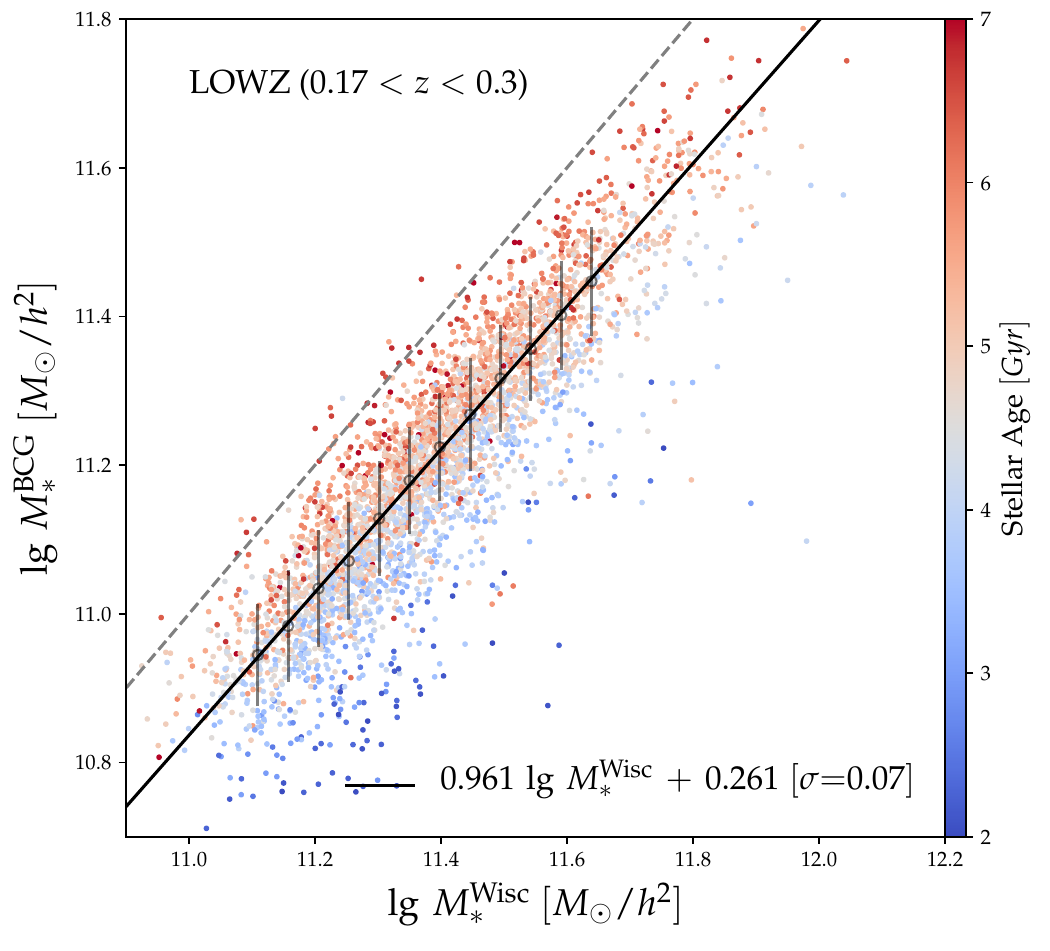}
    \caption{Comparison of the BCG stellar mass estimates derived from
    spectroscopy and photometry, for the redMaPPer BCGs that have spectra from
    the BOSS LOWZ galaxy sample within $z{=}0.17$ and $0.3$. The stellar age of
    each BCG, colour-coded by the colour bar on the right, is estimated from the
    photometric modelling. Black circles are the median stellar mass estimated
    from photometry at fixed spectroscopically-measured stellar mass, with the
    error bars indicating the standard deviation. Black solid line is the linear
    fit to the median relation between the two logarithmic stellar mass
    estimates. Gray dashed line shows the one-to-one line for reference.}
\label{fig:mstar_comparison}
\end{center}
\end{figure}

Adopting the \redmapper{} cluster photometric redshifts for the BCGs, we derive stellar
masses for all $4567$ BCGs by fitting a two-component Simple Stellar Population~(SSP)
template to their SDSS {\it gri} photometry. Following \citet{Maraston2009}, we assume the
dominant stellar population~(97 per cent) to be solar metallicity and include a
secondary~(3 per cent) metal-poor~($Z{=}0.008$) population with the same age.  We utilize
the \texttt{EzGal} software~\citep{Mancone2012} and adopt the \citet{Bruzual2003} SSP
model and the \citet{Chabrier2003} IMF for the fits. We carry
out the fit on extinction-corrected model magnitudes that are scaled to the $i$-band
$c$-model magnitudes. This scaling assumes the outer parts of galaxy profiles are
strictly de Vaucouleurs, which is adequate for isolated elliptical galaxies, but could
underestimate the stellar mass for BCGs by a factor of $2{-}4$~\citep{Bernardi2013,
Kravtsov2018}. Therefore, our stellar mass estimates should be regarded as the mass of the
interior or {\it in-situ} component of the BCGs. The total BCG+ICL~(intra-cluster light)
profiles of the same clusters will be presented in Chen et al.~(in prep).

Figure~\ref{fig:mstar_comparison} shows the comparison between the BCG stellar
mass estimates derived from photometry~($\msbcg$) and from
spectroscopy by \citet{Chen2012}~($M_*^{\mathrm{Wisc}}$). Each galaxy is
colour-coded by the stellar age of the SSP inferred by \texttt{EzGal}, according
to the colour bar on the right.  Our stellar mass estimates are systematically
lower than the \citet{Chen2012} values by $0.1{-}0.15$ dex, indicating by the
black solid line that goes through the median logarithmic $\msbcg$ at
fixed $M_*^{\mathrm{Wisc}}$~(circles). This systematic shift is largely caused by
the different assumptions in the adopted SSP and IMF models, as well as the different
apertures assumed in the two measurements.

\begin{figure}
\begin{center}
    \includegraphics[width=0.48\textwidth]{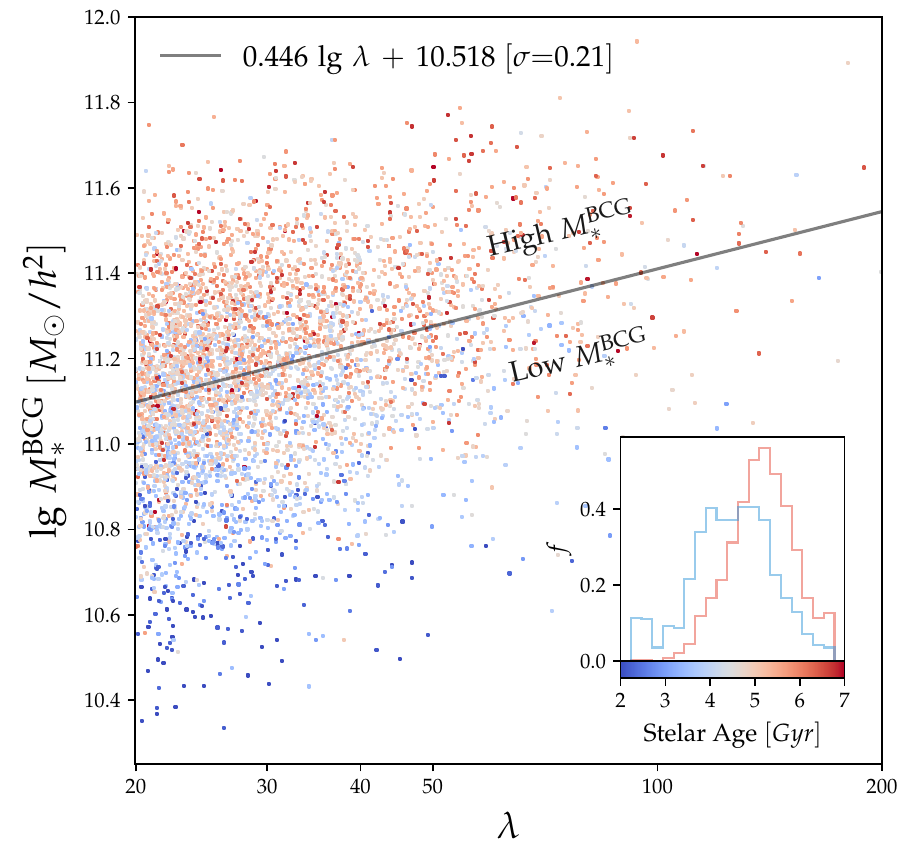}
    \caption{Distribution of clusters on the BCG stellar mass vs. satellite
    richness plane, each colour-coded according to their BCG stellar ages by the
    colour bar underneath the inset panel.  The black line is a linear fit to
    the running median of the logarithmic stellar mass at fixed richness,
    described by the equation on the top left. The cluster sample is split into
    two halves with the same richness distribution but different average BCG
    stellar masses by the black line. The inset panel shows the stellar age
    distribution of the high~(red) and low~(blue) $M_*^{\mathrm{BCG}}$
    subsamples. The average stellar age of the high-$M_*^{\mathrm{BCG}}$ subsample is
    higher than the low-$M_*^{\mathrm{BCG}}$, but the two distributions have significant
    overlap around the median stellar age of the total sample.}
\label{fig:mstarrichness_relation}
\end{center}
\end{figure}

There is also a scatter of ${\sim}0.07$ dex~(errorbars on the circles) between
the \texttt{EzGal} and \citet{Chen2012} masses, mainly due to the differences in
the assumed star formation histories~(SFHs). Our choice of SFH being a single
burst is likely too simplified to describe some of the colour deviations from
predicted by a passively-evolving stellar population of a single age, producing
a degeneracy between the inferred stellar mass and age. Since this extra
age-induced scatter is only one third of the total scatter of $\msbcg$ at fixed
$\lambda$~($0.21$ dex; see Figure~\ref{fig:mstarrichness_relation}), it is
unlikely that the lensing discrepancies~(as we will observe later
in~\S\ref{sec:results}) between the high and low-$\msbcg$ subsamples are caused
by the systematic uncertainties in the assumed SFH. Interestingly,
\citet{Montero-Dorta2017, Niemiec2018} detected significant clustering
difference between the luminous red galaxies split by their SFH, which could be
an indication of the galaxy assembly bias, but at a much higher
redshift~($z{\sim}5$) and for lower mass
systems~(${\sim}8{\times}10^{12}\hmsol$) than our sample.

\subsection{Low and High-$\msbcg$ Cluster Subsamples}
\label{subsec:binning}

Equipped with the homogeneous stellar mass estimates for all $4567$ BCGs in
our cluster sample, we now divide them into subsamples of different average
$\msbcg$ but the same average halo mass. To make sure the signal-to-noise of the
weak lensing measurement of each subsample is high enough for inferring halo
mass and concentration, we do not sub-divide the clusters into smaller bins in
$\lambda$, but introduce a $\lambda$-dependent stellar mass cut to separate the
clusters into two equal-size subsamples with the same distributions of
$\lambda$. Since $\lambda$ is an excellent proxy of halo mass with a tight
scatter~\citep{Rozo2014}, we are hopeful that the two cluster subsamples have similar
average halo masses, as will be seen in ~\S\ref{sec:results}. Similar sliding cuts
across the observable-$\lambda$ plane was also adopted in other cluster assembly bias
studies~\citep{Miyatake2016, Zu2017}.

Figure~\ref{fig:mstarrichness_relation} shows the distribution of clusters on
the $\msbcg$ vs. $\lambda$ plane, with each cluster colour-coded by the stellar
age of its BCG according to the colourbar underneath the inset panel. The gray
line indicates the $\lambda$-dependent stellar mass cut that we adopt to divide
the sample into low vs. high stellar mass halves. For the rest of the paper, we
will refer to the two halves simply as ``low'' and ``high''-$\msbcg$ subsamples,
despite that this is not a uniform stellar mass cut across different $\lambda$.
The scatter of $\msbcg$ at fixed $\lambda$ about the mean $\msbcg$-$\lambda$
relation, described by the parameters in the top left corner, is about $0.21$
dex. If we assume that the additional scatter caused by the stellar age
uncertainties is independent of the true scatter in the SHMR and subtract $0.07$
dex from the quadrature, we arrive at a scatter in the $\msbcg$-$\lambda$
relation of $0.20$ dex, in excellent agreement with spectroscopic observations of the
SHMR from, e.g., \citet{Zu2015, Zu2016}.  The blue and red histograms in the inset
panel of Figure~\ref{fig:mstarrichness_relation}  are the stellar age
distributions of the low and high-$\msbcg$ subsamples, respectively. There is a
substantial overlap in the stellar age distributions of the two populations,
indicating that any observed discrepancy between the two should primarily be
linked to the difference in their average stellar masses instead of age.  After
the split, the difference in the average log-stellar mass between the two
subsamples is about $0.34$ dex~($11.00$ vs.  $11.34$).

\subsection{Cluster Weak Lensing Measurements}
\label{subsec:wl}

We measure the weak gravitational lensing signals for the two cluster subsamples using two
independent measurement methods with separate shear catalogues from two imaging surveys.
We perform the first set of measurements by closely following the method presented in
\citet{Simet2017}. The shear catalogue was derived from SDSS images of
DR8~\citep{Reyes2012} using the re-Gaussianisation algorithm~\citep{Hirata2003} and the
photometric redshifts of the source catalogue were calculated using the Zurich
Extragalactic Bayesian Redshift Analyzer~\citep[ZEBRA][]{Feldmann2006}. Further
characterization of the systematic errors and shear calibrations can be found in
\citet{Mandelbaum2012, Mandelbaum2013}, while the impact of photo-z errors on the weak
lensing measurements can be found in \citet{Nakajima2012}. We refer readers to
\citet{Simet2017} for technical details of this cluster weak lensing measurement based on
the SDSS shear catalogue.

We also perform a second set of weak lensing measurements using the shear
catalogue derived from the DECaLS images of DR8~\citep[Dark Energy Camera Legacy
Survey;][]{Dey2019}, using the photometric redshift estimates from
\citet{Zou2019} via the K-Nearest-Neighbour~(KNN) method.  The sources from the
{\it Tractor} catalogue~\citep{Lang2014} are divided into five morphological
types: Point sources (PSF), round exponential galaxies with a variable radius
(REX), DeVaucouleurs (DEV), Exponential (EXP), and Composite model (COMP).
Sources above 6$\sigma$ detection limit in any stack are kept as candidates. PSF
and REX models are adjusted on individual images convolved by their own PSF
model. Galaxy ellipticities, which are free parameters of the above four REX,
DEV, EXP and COMP models, are estimated by a joint fit on the three optical
$grz$ bands.  We model potential measurement biases with a multiplicative
 and an additive bias~\citep[e.g.,][]{Heymans2012, Miller2013}. The
multiplicative bias comes from the shear measurement and imperfect modeling of PSF size. In order to calibrate our
shear catalogue, we cross-matched the DECaLS DR8 objects with the external shear
measurements, including Canada-France-Hawaii Telescope (CFHT) Stripe
82~\citep{Moraes2014}, Dark Energy Survey~\citep[DES;]{DES2016}, and Kilo-Degree
Survey~\citep[KiDS;]{Hildebrandt2017} objects, and then computed the correction
parameters~\citep{Phriksee2020}. The additive bias is expected to come from
residuals in the anisotropic PSF correction, which depends on galaxy sizes. The
additive bias is subtracted from each galaxy in the catalogue. The same
shear catalogue and photo-z measurements~(but from DECaLS DR3) were used in the
weak lensing analysis of CODEX clusters by \citet{Phriksee2020} and the
intrinsic alignment studies of \citet{Yao2020}, and we refer readers to these
two papers for technical details of the DECaLS shear catalogue and KNN photo-z
errors.

Since DECaLS is roughly 1.5 to 2 magnitudes deeper than SDSS in the r-band, the
effective source number density is higher in DECaLS~($1.8$ arcmin$^{-2}$), about
five times larger than in SDSS~\footnote{We have included the critical surface
density $\Sigma_c$ weighting in the calculation of effective source number
densities.}, while the typical shape noise in DECaLS~($0.23$) is slightly higher
than in SDSS~($0.21$). Therefore, we expect the uncertainties in the DECaLS weak
lensing signal to be smaller by about a factor of two compared to the
SDSS~($\sqrt{5}{\times}(0.21/0.23)$) for the same area coverage~(the effective
source areas of our DECaLS and SDSS measurements are both ${~}9000$ deg$^2$).
For both measurements in SDSS and DECals, we employ the same prescription as
outlined in \citet{Mandelbaum2005, Mandelbaum2013} to apply the ``boost
factor'', to correct for the fact that the lensing signal is diluted by the
inclusion of sources that are physically associated with the lens~(therefore not
really lensed). We derive the uncertainties of both sets of cluster weak lensing
measurements using the Jackknife resampling method, by dividing the sample
footprint into 200 contiguous regions of the same area.

\begin{figure} \begin{center}
    \includegraphics[width=0.48\textwidth]{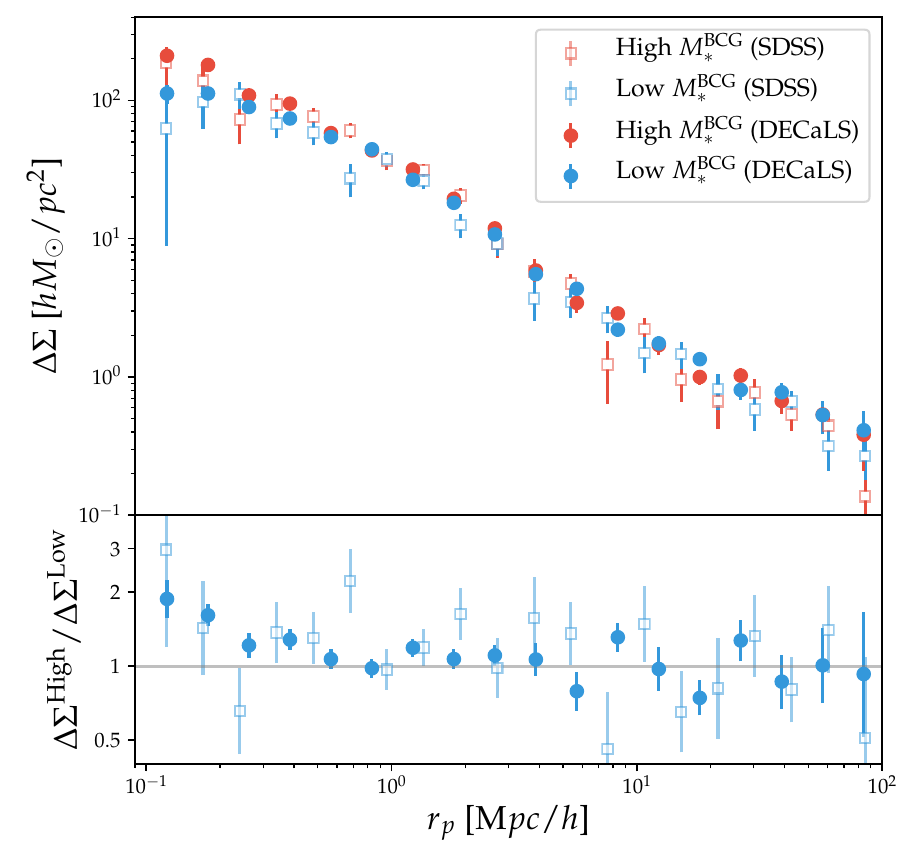} \caption{{\it
    Top:} Surface density contrast profiles of the low~(blue) and high~(red)
    $\msbcg$ subsamples, measured from the SDSS~(open squares with errorbars)
    and DECaLS~(solid circles with errorbars) source catalogues. The smaller
    uncertainties in the DECaLS measurements are mainly due to the
    higher effective number of sources in DECaLS than in SDSS.{\it Bottom:} The ratio
    between the $\ds$ profiles of the high and low-$\msbcg$ subsamples derived
    from SDSS~(open squares) and DECaLS~(filled circles).}
\label{fig:wlsignals} \end{center} \end{figure}

The top panel of Figure~\ref{fig:wlsignals} shows the cluster weak lensing
measurements of the low~(blue) and high~(red) $\msbcg$ subsamples. Open squares
and filled circles are the surface density contrast~($\ds$) profiles measured
from SDSS and DECaLS shear catalogues, respectively.  For each subsample, the
two sets of shear catalogues yield consistent $\ds$ profiles. As expected, the
uncertainties in the DECaLS measurements are smaller than SDSS by a factor of
${\sim}2$, consistent with our expectation.  Focusing on the DECaLS measurements
with smaller errorbars, we find that the two subsamples share similar $\ds$
profiles at projected distances above $1\hmpc$; The two $\ds$ profiles start to
differ below $r_p{\sim}1\hmpc$, with the high-$\msbcg$ subsample showing a much
steeper increase toward smaller scales than the low-$\msbcg$ subsample, which
shows a flattening profile into the inner 100$\hkpc$.  This discrepancy on small
scales can be seen more clearly in the bottom panel of
Figure~\ref{fig:wlsignals}, where we show the ratio of the high and low-$\msbcg$
$\ds$ profiles from SDSS~(open squares) and DECaLS~(filled circles)
measurements. The SDSS measurements display very similar behaviors in the ratio
profile, albeit with much larger uncertainties.

For the small-scale $\ds$ discrepancy, it is very tempting to interpret it as
the manifestation of different halo concentrations at fixed average halo mass,
hence a strong indication that the scatter in the BCG stellar mass is driven by
concentration. However, unlike the X-ray images of clusters that always have
prominent centroids of emission, clusters in the optical do not necessarily have
a dominant BCG in the center. Therefore, a fraction of the BCGs identified by
the optical cluster finder could be incorrect, showing offsets from the X-ray
centres, which are generally believed to be better aligned with the barycentres
of the haloes~\citep{Johnston2007, George2012, Hollowood2019}. Such a miscentring
effect could play an important role in the interpretation of cluster weak
lensing signals on scales up to 1$\hmpc$. Miscentring also adds extra scatter to
the BCG stellar masses by replacing the stellar mass of the true BCG with that
of the incorrect central galaxy. This extra scatter is less problematic because
we can safely assume that the stellar mass of the true BCG is comparable to that
of the miscentred galaxies, because the otherwise dominant true BCG would have
been correctly identified by the centroiding algorithm. To distinguish between
miscentering and concentration effects, and to ascertain whether the average
halo masses are indeed similar, we will derive statistical constraints on halo
mass, miscentring, and concentration by modelling the $\ds$ profiles measured
from DECaLS imaging in the next section.

\section{Bayesian Inference of Cluster Properties}
\label{sec:bayes}

\subsection{Theoretical Modelling of $\ds$}
\label{subsec:ds}

We adopt a Bayesian framework for inferring the cluster density profiles from
weak lensing signals. The surface density contrast profile $\ds$ can
be computed as
\begin{equation}
    \ds(r_p) = \overline{\Sigma}({<}r_p) - \Sigma(r_p),
\end{equation}
where $\overline{\Sigma}({<}r_p)$ and $\Sigma(r_p)$ are the average surface
matter density interior to and at radius $r_p$, respectively.  In the absence of
miscentring, $\Sigma(r_p)$ can be directly computed from the 3D isotropic
halo-mass cross-correlation function $\xihm(r)$,
\begin{equation}
    \Sigma(r_p) = \rho_{m} \int_{-\infty}^{+\infty}
\xi_{\mathrm{hm}}(r_p,\,r_{\pi})\;\mathrm{d} r_{\pi},
\end{equation}
where $\rho_{m}$ is the mean density of the Universe. In practice, we
ignore the effects of the broad lensing window function, and
use $\pm\,100\,\hmpc$ for the integration limit.

To model the miscentering effect, we assume that the fraction of BCGs
miscentred is $f_{\mathrm{off}}$ and their offsets $r_{\mathrm{off}}$ from the
true centres follow a shape-2 Gamma distribution $p(r_{\mathrm{off}})$ with a
characteristic offset $\sigma_{\mathrm{off}}$,
\begin{equation}
    p(r_{\mathrm{off}}) =  \frac{r_{\mathrm{off}}}{\sigma_{\mathrm{off}}^2}
    \exp\left(-\frac{r_{\mathrm{off}}}{\sigma_{\mathrm{off}}}\right).
\end{equation}
The observed surface matter density in the presence of miscentring is thus
\begin{equation}
    \Sigma^{\mathrm{obs}}(r_p) =
    f_{\mathrm{off}}\,\Sigma^{\mathrm{off}}(r_p)\; +\; (1 -
    f_{\mathrm{off}})\,\Sigma(r_p),
\end{equation}
where
\begin{equation}
    \Sigma^{\mathrm{off}}(r_p) = \frac{1}{2\pi}
    \int_0^{\infty}\!\!\!\!\!\mathrm{d} r_{\mathrm{off}}\, p(r_{\mathrm{off}})
    \int_0^{2\pi} \!\!\!\!\!\mathrm{d}\theta\,\Sigma\left(\sqrt{r_p^2 + r_{\mathrm{off}}^2 - 2 r_p
    r_{\mathrm{off}} \cos\theta }\right).
\end{equation}

Finally, to calculate $\Sigma(r_p)$, we need to build an accurate model for
$\xi_{\mathrm{hm}}$ on radial scales between 0.1$\hmpc$ and ${\sim}100\hmpc$.
We adopt the two-component model of $\xi_{\mathrm{hm}}$ developed by
\citet{Zu2014}~\citep[a modified version proposed by][]{Hayashi2008},
\begin{eqnarray}
\xihm(r) &=&   \left\{
\begin{array}{ll}
 \xi_\mrm{1h} & \quad\mbox{if $\xi_\mrm{1h} \geqslant \xi_\mrm{2h}$ },\nonumber\\
 \xi_\mrm{2h} & \quad\mbox{if $\xi_\mrm{1h} < \xi_\mrm{2h}$ },\nonumber
\end{array}
\right.\\
\xi_\mrm{1h} &=& \frac{\rho_\mrm{NFW}(r|M_h)}{\rho_\mrm{m}} - 1, \nonumber\\
\xi_\mrm{2h} &=& b \; \ximm.
\label{eqn:xihm}
\end{eqnarray}
Here $\xi_\mrm{1h}$ and $\xi_\mrm{2h}$ are the so-called ``1-halo'' and
``2-halo'' terms in the halo model~\citep{Cooray2002}, $\rho_\mrm{NFW}(r|M_h,
c)$ is the NFW density profile of halo mass $M_h$ and concentration $c$, $b$ is
the average halo bias, and $\ximm$ is the non-linear matter-matter
auto-correlation function predicted at {\it Planck}
cosmology~\citep{Takahashi2012}.  \citet{Zu2014} found that this simple model
provides an adequate description of the halo-matter cross-correlation measured
from simulations at the level of a few per cent on scales of our concern~(i.e.,
below the halo radius). For massive haloes, several studies found that the
average density profile in the inner region of haloes deviates from the NFW
shape and the Einasto profile is more accurate~\citep{Dutton2014, Klypin2016}.
However, since we are only fitting to scales above $0.1\hmpc$, the difference
between NFW and Einasto should be negligible. We have also ignored the extra
lensing effect caused by the stellar mass of the BCGs, which has negligible
contribution on scales above $0.1\hmpc$.
The uncertainties of the predicted $\ds$ can be as large as $10\%$
around the transition between the 1-halo and 2-halo scales~\citep[$2{-}4\hmpc$;
see the figure 5 of][]{Zu2014}, but we are only concerned with the
measurement of halo mass and concentration, which are inferred primarily from
scales below the transition scale, and the large-scale bias, which is estimated
from scales above $10\hmpc$. Additional, our statistical uncertainties of weak lensing at
those scales is around ${~}10\%$. Therefore, the impact of relatively large uncertainties
in the transitional regime on our conclusions should be negligible. For future
modelling of cluster weak lensing signals of per cent level uncertainties, the
transitional behavior can be potentially improved by using the scheme proposed by
\citet{Garcia2020} and emulators developed by \citet{Salcedo2020}.

\subsection{Model Priors and Likelihood}
\label{subsec:prior}

For each cluster subsample, we have five parameters for modelling the $\ds$
profile: three of them for describing $\xihm$~($M_h$, $c$,  $b$) and two for
miscentring~($f_{\mathrm{off}}$, $\sigma_{\mathrm{off}}$). The miscentring
effect is highly degenerate with halo concentration if $\sigma_{\mathrm{off}}$
is allowed to vary arbitrarily. To mitigate such degeneracy in our constraint,
we apply the results from the state-of-the-art calibration of the \redmapper{}
cluster miscentring from \citet{Zhang2019} as our priors for modelling the
miscentring. Using the X-ray observations from {\it Chandra}, their constraints
on the average offset and the average miscentring fraction are~(listed in their
table 1) $\avg{\sigma_{\mathrm{off}}}{=}0.18\pm0.02\hmpc$ and
$\avg{f_{\mathrm{off}}}{=}0.3\pm0.04$, respectively.

However, the calibration in \citet{Zhang2019} was derived for the overall SDSS
\redmapper{} sample, which is not directly applicable to our cluster subsamples
split by $\msbcg$. For example, for our high-$\msbcg$ subsample it is reasonable
to expect the miscentring fraction $f_{\mathrm{off}}$ to be lower than $0.3$,
because more massive \redmapper{} BCGs are more likely to be the correct central
galaxies, and vice versa for the low-$\msbcg$ subsample. To correctly make use
of the \citet{Zhang2019} constraints in our analysis, we fit the $\ds$ profiles
of the low and high-$\msbcg$ subsamples jointly as a single data vector with ten
model parameters~(five for each subsample). During the fit, we derive the
average miscentring parameters $\avg{\sigma_{\mathrm{off}}}$ and
$\avg{f_{\mathrm{off}}}$ for the overall cluster sample in each likelihood
calculation, and apply the \citet{Zhang2019} constraints as Gaussian priors on
those two quantities, so that $\avg{\sigma}_{\mathrm{off}}\sim
\mathcal{N}(0.18\hmpc, 0.02^2)$ and $\avg{f}_{\mathrm{off}}\sim \mathcal{N}(0.3,
0.04^2)$, respectively. In this way, we implement the miscentring priors in our
analysis self-consistently and correctly take into account the covariance
between the two sets of $\ds$ parameters during the fit.

The concentration of a typical halo with $\lg\,M_h=2\times10^{14}\hmsol$ is
predicted to be roughly five by the $\lcdm$ model at {\it Planck} cosmology, so
we place a broad Gaussian prior of the average concentration of each subsample
as $c{\sim}\mathcal{N}(5, 1.5^2)$. Note that the average concentration of our
cluster sample may be significantly higher than five, due to the increasing
scatter of the mass-richness relation toward the low-$\lambda$
end~\citep{Murata2018}. Since the $c$-$m$ relation increases rapidly toward the
low-mass end, the progressively large scatter will introduce a large number of
low-mass haloes that usually have significantly higher concentrations into our cluster
sample. Nonetheless, this prior tends to drive the concentration values of the
two subsamples closer during the fit, and is thus a conservative choice given
our purpose of looking for discrepancies in $c$. We assume flat priors for the
other two parameters of $\xihm$.

We assume a Gaussian likelihood model and compute the likelihoods by comparing
the predicted $\ds$ to the DECaLS measurements on scales between $0.1\hmpc$ and
$20.0\hmpc$.  To infer the joint posterior distribution of the ten parameters,
we employ the affine invariant Markov Chain Monte Carlo~(MCMC) ensemble sampler
\texttt{emcee}~\citep{Foreman-Mackey2013}. We run the MCMC sampler for
$2,000,000$ steps for each analysis to ensure its convergence, and derive the
posterior constraints after a burn-in period of $500,000$ steps. The median
values and the 68 per cent confidence limits of the 1D posterior constraints are
listed in Table~\ref{tab:constraints}, with one row for each cluster subsample.

\section{Constraints on Halo Properties}
\label{sec:results}

\begin{table}
\renewcommand*{\arraystretch}{1.3}
\centering \caption{Posterior constraints of the model parameters for the two subsamples. The uncertainties are the $68\%$ confidence regions derived from the 1D posterior probability distributions.}
\begin{tabular}{cccccc}
\hline
\hline
    Subsample & \hspace{-10pt} $\lg\,M_h$ &\hspace{-10pt} $c$ &\hspace{-10pt} $b$ &\hspace{-10pt} $\sigma_{\mathrm{off}}$ &\hspace{-10pt} $f_{\mathrm{off}}$ \\
\hline
    Low-$\msbcg$ &\hspace{-10pt}  $14.24_{-0.02}^{+0.02}$ &\hspace{-10pt} $5.87_{-0.60}^{+0.77}$ &\hspace{-10pt} $2.91_{-0.27}^{+0.26}$ &\hspace{-10pt} $0.23_{-0.02}^{+0.02}$ &\hspace{-10pt} $0.37_{-0.06}^{+0.07}$ \\
    High-$\msbcg$ &\hspace{-10pt}  $14.24_{-0.02}^{+0.02}$ &\hspace{-10pt} $6.95_{-0.66}^{+0.78}$ &\hspace{-10pt} $2.59_{-0.21}^{+0.21}$ &\hspace{-10pt} $0.21_{-0.03}^{+0.02}$ &\hspace{-10pt} $0.20_{-0.07}^{+0.07}$ \\
\hline
\end{tabular}
\label{tab:constraints}
\end{table}

\begin{figure*}
\begin{center}
    \includegraphics[width=0.96\textwidth]{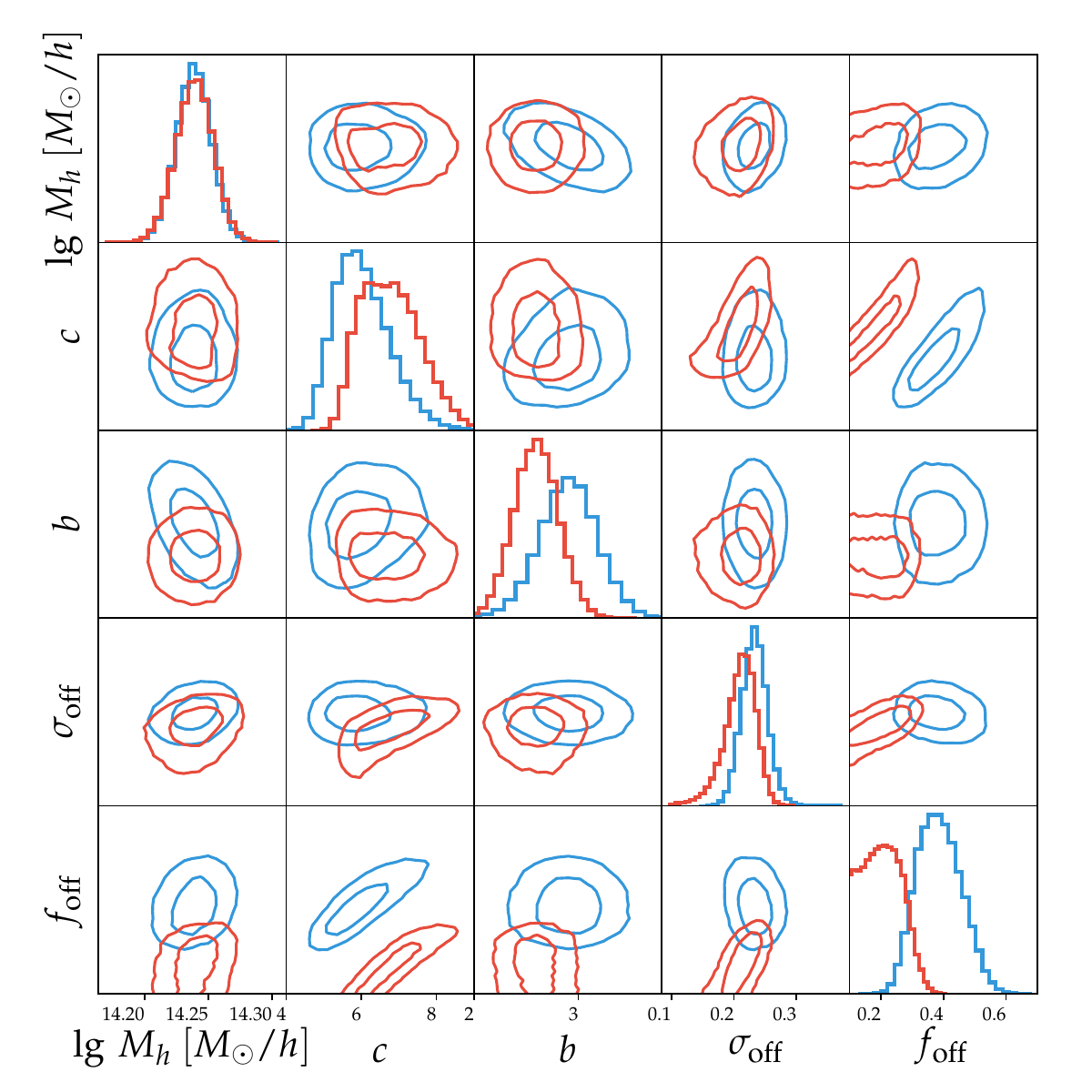}
    \caption{Posterior constraints from the modelling of $\ds$ of the low~(blue) and
    high~(red) $\msbcg$ subsamples using the DECaLS imaging data. Each contour
    set in the off-diagonal panels indicate
    the $50\%$~(inner contour) and $90\%$~(outer contour) confidence regions of the
    matching parameter pair, and each
    histogram in the diagonal panels are the 1D marginalised posterior distribution of the
    respective parameter.}
\label{fig:posterior}
\end{center}
\end{figure*}

The derived posterior constraints of our model parameters for the low~(blue) and
high~(red) $\msbcg$ subsamples are displayed in Figure~\ref{fig:posterior}. The diagonal
panels show the 1D marginalised posterior distributions of each of the five parameters,
and the contours in the off-diagonal panels are the $50\%$ and $90\%$ confidence regions
for each of the parameter combinations within each subsample. Note that we do
not show the parameter covariances between the two subsamples because they are
either very weak~($\lg\,M_h$, $c$, $b$) or can be entirely explained by the
priors on the overall miscentring~(${f_{\mathrm{off}}}$,
${\sigma_{\mathrm{off}}}$).

\begin{figure*}
\begin{center}
    \includegraphics[width=0.96\textwidth]{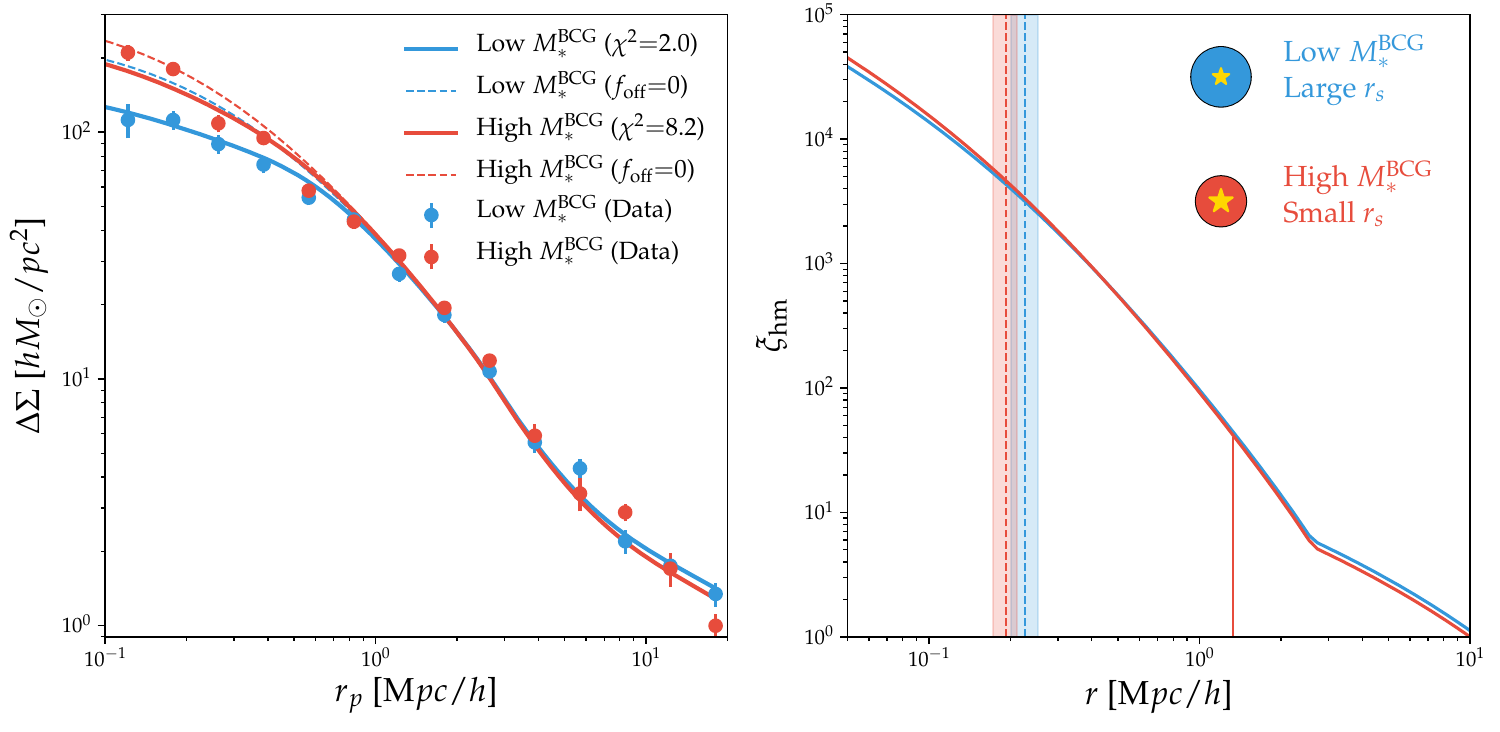} \caption{{\it
    Left panel}: Comparison between the $\ds$ predicted by the posterior mean
    models~(thick solid curves) and the DECaLS measurements~(circles with
    errorbars), for the low~(blue) and high~(red) $\msbcg$ subsamples,
    respectively. The $\chi^2$ values of the two posterior mean models are shown in the parentheses. Thin dashed curves of the matching colours indicate the model
    predictions in the absence of miscentring. {\it Right panel}: Comparison of
    the halo-matter cross-correlation functions between the low~(blue) and
    high~(red) $\msbcg$ subsamples, inferred from the posterior mean models of
    the $\ds$ analysis~(Equation~\ref{eqn:xihm}).  Blue and red vertical lines underneath the $\xihm$
    curves at $r{\simeq}1.33\hmpc$ indicate the best-fitting $r_{200m}$ of the
    low and high-$\msbcg$ subsample, respectively, while the blue and red shaded
    vertical bands indicate the 1-$\sigma$ ranges of characteristic radii $r_s$
    of the two corresponding best-fitting NFW profiles. The $r_{200m}$ lines are
    almost indistinguishable
    because the inferred halo masses are similar, while the two $r_s$ values differ by ${\sim}20\%$. The two filled circles illustrate the size difference between the two $r_s$ in linear scale, with the gold stars in the center representing the different average BCG stellar masses.}
\label{fig:wlmodels}
\end{center}
\end{figure*}

The halo mass constraints of the two subsamples are almost exactly the
same~($\lg\,M_h{=}14.24{\pm}0.02$), confirming our expectations that the two
subsamples have the same average halo mass.
Interestingly, the model yields a
${\sim}12\%$ higher bias for the low-$\msbcg$ subsample~($2.91{\pm}0.27$) than the
high-$\msbcg$ one~($2.59{\pm}0.21$). However, the biases are less well
constrained due to the relatively large uncertainties of $\ds$ on scales above
$10\hmpc$. We will examine the biases more closely in ~\S\ref{sec:cross} using
cross-correlations with galaxies.  The characteristic offset of the
high-$\msbcg$ clusters~($0.21{\pm}0.02\hmpc$) is slightly smaller compared to
the low-$\msbcg$ subsample~($0.23{\pm}0.02\hmpc$), but their inferred miscentring
fraction~($0.20{\pm}0.07$) is lower than that of
the low-$\msbcg$ clusters~($0.37{\pm}0.07$) by ${\sim}1.5\sigma$. This indicates that the
high-$\msbcg$ clusters are indeed better centered than their low-$\msbcg$
counterparts, and some of the small-scale discrepancies between the two $\ds$
profiles are due to a stronger miscentring effect in the low-$\msbcg$ subsample.

As expected from~\S\ref{subsec:prior}, the concentration values inferred for
both subsamples are higher than five, the predicted value for a halo with
$\lg\,M_h{\simeq}14.2$ in $\lcdm$, likely due to the contribution from a large
number of intrinsically low-mass haloes with much higher concentration. The average
concentration of the overall cluster sample is ${\sim}5{-}7$, consistent with the
concentration measurements from \citet{Shan2017} and \citet{Simet2017}.

After marginalising over the uncertainties in the miscentring, the two 1D
posterior constraints on the halo concentrations exhibit a statistically
significant difference~(${>}1\sigma$), with the low-$\msbcg$ clusters having a
lower concentration~($5.87_{-0.60}^{+0.77}$) than the high-$\msbcg$
counterparts~($6.95_{-0.66}^{+0.78}$). For the discrepancy in the small-scale
weak lensing signals to be entirely attributed to the difference in miscentring,
the ratio of two miscentering fractions has to exceed ${\sim}4{:}1$, which is
mathematically plausible but highly unlikely.  The concentration of the
low-$\msbcg$ subsample is roughly consistent with the input Gaussian prior on
concentration.  For the high-$\msbcg$ subsample, it requires a miscentring
fraction $f_{\mathrm{off}}$ of almost zero to reach the average concentration of
the low-$\msbcg$ subsample~($5.87$), and a negative $f_{\mathrm{off}}$~(i.e.,
unphysical) to be consistent with the prior.

Therefore, we believe the inferred high concentration of the high-$\msbcg$
clusters is caused by the preference of massive BCGs to live in more
concentrated haloes.  Given that the two subsamples have the same average halo
mass but differ significantly in their BCG stellar mass, our result here
suggests that the dark matter concentration is indeed strongly tied to the
stellar mass of the central galaxy in massive haloes. Since the concentration of
a halo was set at the very early stage of halo formation, it is more likely that
concentration drives the scatter in the SHMR of clusters, rather
than vice versa.

Linking the discrepancies in both concentration and bias between the two
subsamples, our results indicate that the low-$\msbcg$ clusters are less
concentrated on small scales but more clustered on large-scales than their
high-$\msbcg$ counterparts of the same halo mass. This observed secondary
dependence of halo bias on concentration at fixed halo mass, if confirmed by
more accurate measurements of bias, provides direct evidence for the existence
of cluster assembly bias~\citep{Gao2005, Jing2007}. We will return to the
assembly bias discussion in~\S\ref{sec:cross}.

\begin{figure*}
\begin{center}
    \includegraphics[width=0.96\textwidth]{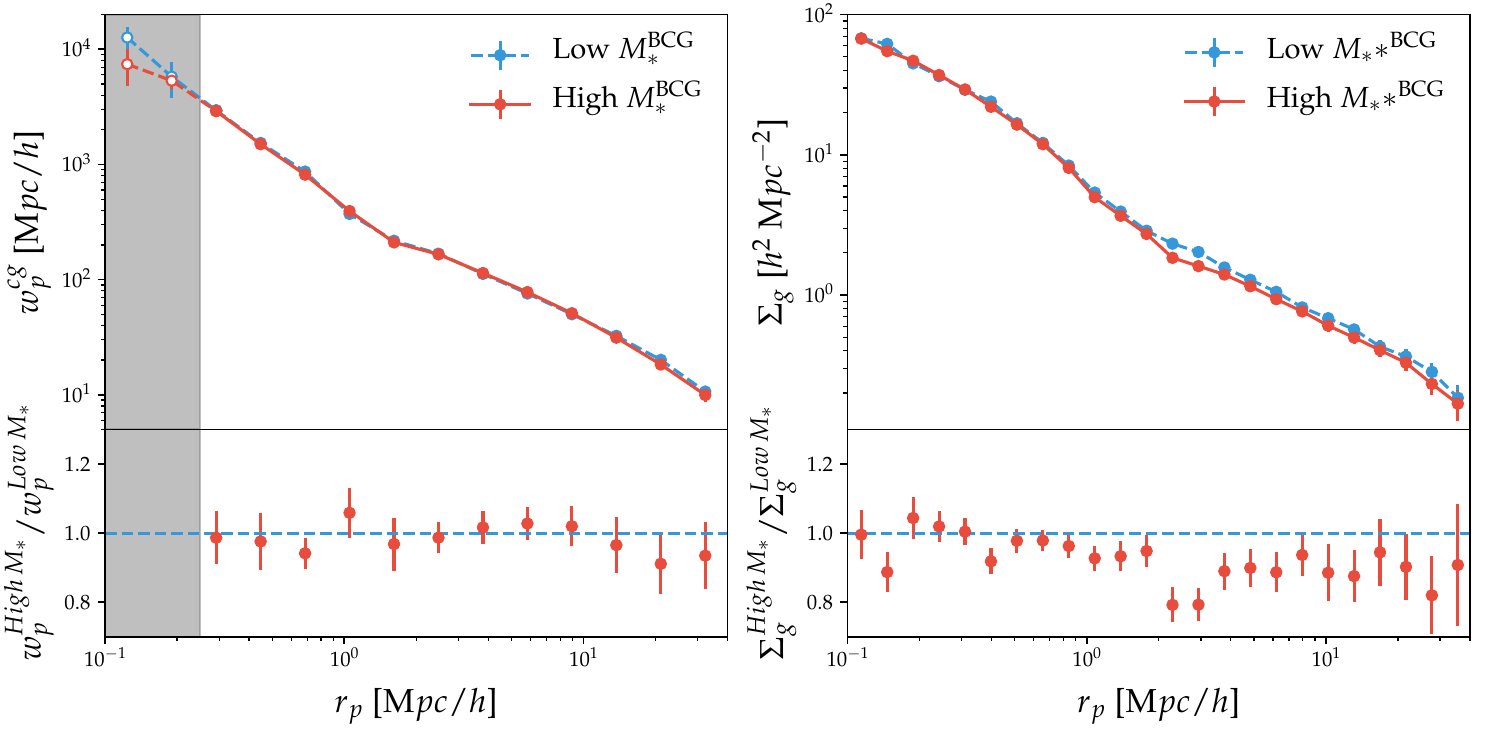} \caption{{\it
    Left panel}: Comparison of the projected correlation functions of the
    low~(blue) and high~(red) $\msbcg$ subsamples with the BOSS LOWZ
    spectroscopic galaxy sample in the upper subpanel, with the ratio of the two
    shown in the bottom subpanel. The gray shaded region on the left indicates
    the distance scales that are affected by the fibre collision in BOSS. {\it
    Right panel}: Similar to the left panel, but for the excess galaxy surface
    number density profile calculated from the SDSS DR8 imaging. In both panels, the two
    profiles are consistent on scales below $1\hmpc$, confirming that the satellite
    richnesses of the two subsamples are the same; The low-$\msbcg$ subsample has a
    slightly higher clustering bias than the high-$\msbcg$ subsample on scales above
    $10\hmpc$, consistent with the weak lensing analysis.}
\label{fig:wpsignals}
\end{center}
\end{figure*}

Our inferred concentration-$\msbcg$ connection and the evidence of halo assembly
bias point to a simple yet remarkable picture of cluster formation across
distance scales of three orders of magnitudes, from the stellar mass within tens
of $k$pc, to dark matter density within the scale radius of hundreds of $k$pc,
and finally to the large-scale overdensity of cluster haloes on several tens of
Mpc. In this picture, at fixed cluster mass and richness, the highly
concentrated haloes host more massive galaxies in the center, but
preferentially live in lower-density large-scale environments.

The left panel of Figure~\ref{fig:wlmodels} compares the $\ds$ predicted by the
posterior mean models~(thick solid curves) to the DECaLS observations~(circles
with errorbars), for the low~(blue) and high~(red) $\msbcg$ subsamples.  The
high-$\msbcg$ observations are slightly higher than predicted on scales below
$0.2\hmpc$, probably due to some unknown deviations from our simple miscentring
model  or underestimated systematic uncertainties in the
small-scale weak lensing measurements. Overall, the model predictions provide
a very good overall description to the data points on all scales for both
subsamples. We also show the two model predictions assuming there is no
miscentring with the thin dashed curves, by setting $f_{\mathrm{off}}{=}0$ while
keeping all other parameters fixed to the posterior mean values. Clearly,
miscentring incurs a much larger suppression in the weak lensing signals of the
low-$\msbcg$ subsample~(blue dashed thin curve) than that of the high-$\msbcg$
subsample~(red dashed thin curve) on smaller scales. Nevertheless, the model
predicts that approximately half of the observed small-scale discrepancy in
$\ds$ is induced by the intrinsic difference in their average halo
concentrations.

On the right panel of Figure~\ref{fig:wlmodels}, we illustrate the two
corresponding $\xihm$ profiles predicted by the posterior mean models for the
low~(blue) and high~(red) $\msbcg$ subsamples. The corresponding vertical lines
underneath the $\xihm$ curves indicate the predicted $r_{200m}$, which are
indistinguishable between the two subsamples. The blue and red shaded vertical
bands indicate the $1-\sigma$ ranges of the two characteristic scale radii $r_s$
of the low and high-$\msbcg$ subsamples, respectively. In a more visually
appealing way, the blue and red filled circles in the top right corner compare
the relative sizes of the two $r_s$ values in linear proportion, and the stars in the
center roughly illustrate the different average stellar masses of the central
galaxies --- at fixed halo mass, more concentrated clusters have more massive
central galaxies.

Therefore, our weak lensing analysis suggests that the two cluster subsamples, divided by
their BCG stellar masses, have very similar average halo masses, but likely differ in
their concentration parameters by ${\sim}18\%$. The simplest explanation is that, the halo
concentration parameter is the key driver in setting the stellar mass of the
BCGs in massive clusters. Intriguingly, the concentration-stellar mass relation
at fixed halo mass is also directly linked to the halo bias, which we further
investigate in the next section using cluster-galaxy cross-correlations.

\section{Cluster Assembly Bias from Cross-Correlations with Galaxies}
\label{sec:cross}

From the weak lensing analysis, we find that the two cluster subsamples split by
$\msbcg$ have very similar average halo masses, but differ in their
concentrations after marginalising over the degeneracy between concentration and
miscentring effects. Additionally, the weak lensing analysis also suggests that
the two subsamples differ in halo bias, a possible manifestation of the
cluster assembly bias effect that has yet to be detected~\citep{Miyatake2016,
Zu2017}.

To further investigate whether the two subsamples have different halo bias, we
measure their projected cross-correlation functions with BOSS LOWZ
spectroscopic~($w_p^{cg}$) and SDSS photometric galaxies~($\Sigma_{g}$), shown
in the left and right panels of Figure~\ref{fig:wpsignals}, respectively. In
each panel, blue and red circles with errorbars are the measurements for the low
and high-$\msbcg$ subsamples, respectively, and circles with errorbars in the
bottom subpanel show the ratios between the high and low-$\msbcg$ measurements.
The error matrices are estimated from the same Jackknife resampling technique as in the cluster weak lensing measurements.
The gray shaded band in the left panel indicates the scales affected by the
fibre collision in BOSS. For the sake of brevity, we directly present the
measurement results in this paper without the technical details, as we are only
concerned with the relative difference between the large-scale clustering of the
two subsamples. We faithfully follow ~\citet{Zu2020} for the LOWZ galaxy sample
selection and the $w_p$ calculation, while the SDSS imaging galaxies and the
calculation of surface number density profile $\Sigma_{g}$ can be found in
\citet{More2016}.

Clearly, the two sets of measurements in Figure~\ref{fig:wpsignals} are consistent with
our results from the weak lensing analysis.  That is, the low-$\msbcg$ cluster subsample
has a ${\sim}10\%$ higher large-scale clustering than their high-$\msbcg$ counterparts on
scales larger than $10\hmpc$. The bias discrepancy is around 1 $\sigma$ for
$w_p^{cg}$, and close to 1.5 $\sigma$ for $\Sigma_{g}$. The two projected
spectroscopic correlation functions $w_p^{cg}$ start to differ at
${\sim}10\hmpc$, while the two photometric galaxy number density profiles
$\Sigma_g$ bifurcate on much smaller scales~($2\hmpc$), probably because the
photometric signal is dominated by galaxies that are much fainter than the LOWZ
galaxies and have a different scale--dependent assembly bias. On scales below
$1\hmpc$, the two subsample shows very similar galaxy correlation signals, which
is by design because they have the same satellite richness and similar weak
lensing halo masses.

It is interesting if the cross-correlations with galaxies shown in
Figure~\ref{fig:wpsignals} would show any evidence of difference in the galaxy
concentrations on small scales. For $w_p^{cg}$, unfortunately the fibre collision scale is
comparable to the $r_s$ inferred from weak lensing, preventing any meaningful measurement
of galaxy concentration. The cluster cross-correlations with photometric galaxies are free
of fibre collision, and indeed exhibit a weak evidence that the satellite galaxies in the
high-$\msbcg$ clusters are slightly more concentrated than those in the low-$\msbcg$
systems on scales below $1 \hmpc$. However, the uncertainties are too large to make a more
concrete statistical statement.

To summarise, the cluster-galaxy cross-correlation functions provide further support of our
conclusion from the weak lensing analysis, that the large-scale bias of the
low-$\msbcg$, low-$c$ clusters is $10\%$ higher than that of their high-$\msbcg$, high-$c$
counterparts with the same average halo mass. The {\it physical} connection of
the three key cluster properties across scales of almost three orders of
magnitudes, the stellar content within $20{-}30\hkpc$, the dark matter density
within ${\sim}200\hkpc$, and the cosmic overdensity on scales above $10\hmpc$,
is remarkable, revealing a surprisingly elegant picture of the co-evolution of
massive dark matter haloes and their central galaxies amid the hierarchical structure
formation of our Universe.

\section{A Conundrum: Does Halo Mass Depend on $M_*^{\mathrm{BCG}}$ at fixed $\lambda$?}
\label{sec:puzzle}

Before concluding our work, we would like to briefly discuss an apparent halo
mass conundrum in our result --- the average halo masses of the low and
high-$\msbcg$ cluster subsamples are indistinguishable, despite the SHMR
predicts that the average halo mass of the high-$\msbcg$ clusters should be
higher~\citep[e.g., see figure 11 of][]{Zu2015}. In particular, since there is a
${\sim}0.25$ dex scatter between our mass proxy $\lambda$ and the true halo
mass, the halo mass distribution at fixed $\lambda$ should be fairly broad,
especially at the low-$\lambda$ end~\citep[see figure 7 of ][]{Murata2018}.
Naively, one might expect that the SHMR would still operate at fixed $\lambda$,
so that the more massive BCGs would preferentially live in high-mass haloes than
the less massive systems, creating a mass discrepancy around $0.2{-}0.3$ dex
between the two subsamples based on their $0.34$ dex difference in average
stellar mass. However, our weak lensing analysis indicates that the mass
discrepancy, if exists at all, should be smaller than $0.04$ dex, i.e., $10\%$.

In addition, the large-scale bias of the low-$\msbcg$ subsample is $10\%$ higher than that
of the high-$\msbcg$ subsample, further indicating that the halo mass of the
high-$\msbcg$ subsample is unlikely larger than that of the low-$\msbcg$ one.
The reason is as follows. The average halo bias increases steeply with halo mass
above the characteristic nonlinear mass scale, and for haloes of
$\lg\,M_h{\simeq}14.2$ at the mean redshift of our sample, the slope
$\mathrm{d}\ln\,b/ \mathrm{d}\ln\,M_h$ is about $0.42$.  Therefore, a
$0.2{-}0.3$ dex halo mass enhancement in the high-$\msbcg$ subsample
would lead to a $20{-}30\%$ higher bias of the high-$\msbcg$ subsample than the
low-$\msbcg$ one~(in the absence of halo assembly bias), yet we observe a $10\%$
lower bias of the high-$\msbcg$ subsample, i.e., a $30{-}40\%$ bias inversion.

This strong bias inversion can not be entirely explained by halo assembly bias, if the
halo mass of the high-$\msbcg$ clusters were indeed $0.2{-}0.3$ dex higher than their
low-$\msbcg$ counterparts. \citet{Jing2007} predicted that the bias of the $20\%$ most
concentrated haloes is $20\%$ lower than the $20\%$ least concentrated haloes in the
cluster regime.  Therefore, even if we assume an extreme scenario in which the low and
high-$\msbcg$ clusters correspond to the $20\%$ least and most concentrated haloes,
respectively, the halo assembly bias effect would not be able to resolve the large bias
inversion we observed in the data.

This is very intriguing. One possible explanation is that the stellar mass of
the BCG correlates strongly with the satellite richness at any given halo mass,
so that the halo mass distribution at fixed $\lambda$ is non-trivially related
to $\msbcg$. Evidence of such a correlation was recently suggested by
~\citet{To2020}. In a future work~(Zu et al. in prep), we will explore whether
the correlation has to be positive or negative, and what halo mass dependence of
the correlation is required to solve the conundrum we observed here.

Finally, we emphasize that the main conclusion of this paper, which is derived
from the observed {\it average} properties of the two cluster subsamples, is
independent of the solution to such a halo mass conundrum, as long as the two
share the same weak lensing halo mass.

\section{Conclusion}
\label{sec:conc}

In this paper, we have investigated the origin of the scatter in the SHMR of
SDSS \redmapper{} clusters by examining the weak gravitational lensing signals
of two cluster subsamples split by their BCG stellar mass at fixed richness. To
overcome the spectroscopic incompleteness of the BCG sample, we derived a
homogeneous measurement of the BCG stellar masses from the SDSS {\it gri}
photometry using the accurate cluster photometric redshifts estimated by the
\redmapper{}.  Finally, using the shear catalogue derived from the DECaLS and
SDSS imaging, we obtained accurate cluster weak lensing profiles $\ds$ for the
two subsamples, which exhibit strong discrepancies on small scales.

To interpret the discrepancies in $\ds$, we modelled the DECaLS $\ds$ profiles
using the halo-matter cross-correlation prescription of \citet{Zu2014}~\citep[a
variant modified from][]{Hayashi2008}, and carefully took into account the
cluster miscentring calibrations from \citet{Zhang2019}. After marginalising
over the uncertainties in miscentring, we found that the two subsamples of
clusters have almost the same average halo mass~$\lg\,M_h{=}14.24{\pm}0.02$,
but different average values of halo concentration. In particular, the
low-$\msbcg$ clusters are on average less
concentrated~($c{=}5.87_{-0.60}^{+0.77}$) than the high-$\msbcg$
systems~($6.95_{-0.66}^{+0.78}$). Our results provide direct evidence that the
scatter in the stellar mass of central galaxies at fixed halo mass is strongly
tied to the dark matter concentration of clusters.

Furthermore, we found that the low-$\msbcg$, low-$c$ clusters are more clustered
on scales above $10\hmpc$ than their high-$\msbcg$, high-$c$ counterparts from
the weak lensing analysis. This finding is corroborated by our
cluster-galaxy cross-correlation measurements using both the SDSS LOWZ
spectroscopic and DR8 photometric galaxy samples. We interpret this observed
dependence of large-scale bias on halo concentration at fixed cluster mass as a
possible detection of the halo assembly bias effect, consistent with the
expectations from the $\lcdm$ simulations~\citep{Gao2005, Jing2007, Zu2017}.

Combining our two key findings, we infer that there likely exists a
{\it physical} connection among the three key cluster properties across scales
of almost three orders of magnitudes, the stellar mass within $20{-}30\hkpc$,
the dark matter mass within ${\sim}200\hkpc$, and the cosmic overdensity on
scales above $10\hmpc$, for clusters of the same average halo mass. This
remarkable connection links the origin the scatter in the cluster SHMR to the
physics of halo assembly bias, thereby revealing a powerfully simple picture
of the massive galaxy formation within the hierarchical structure formation in
our Universe.

The main sources of systematic uncertainties in our
analysis are the modelling of miscentring effect and the accuracy of stellar
mass estimates. The miscentring modelling in our analysis can be immediately
improved by applying the \citet{Zhang2019} calibration to the two cluster
subsamples separately. However, the current number of clusters with
high-resolution X-ray observations is rather limited, and the miscentring
calibration will benefit greatly from the upcoming X-ray surveys like the
eROSITA~\citep{Merloni2012} and improved cluster centroiding algorithms.
Meanwhile, the stellar mass estimates can be significantly improved with the
upcoming observations by DESI~\citep{DESI2016} and PFS~\citep{Takada2014}, which
will eliminate the spectroscopic incompleteness of the low-mass BCGs with bluer
colours.

Looking forward, current deep imaging surveys like the DES~\citep{DES2016},
HSC~\citep{HSC2018}, and LSST~\citep[Rubin;][]{LSST2019} on the ground and future
space missions like the {\it WFIRST}~\citep[Roman;][]{WFIRST2013}, {\it
Euclid}~\citep{Laureijs2011}, and {\it CSST}~\citep[Chinese Space Station
Telescope;][]{CSST2019} will provide superb photometry for the measurement of
stellar mass profiles of massive centrals galaxies~\citep{Huang2020} and the
weak gravitational lensing signals of their host dark matter
halos~\citep{Mandelbaum2018}. Finally, our inferred picture of cluster
formation connects the structure formation across scales of three orders of
magnitudes, thereby providing a stringent test of the theories of co-evolution
between massive clusters and their brightest central galaxies~\citep{Naab2017,
Pillepich2018}.

\section*{Acknowledgements}

We thank the anonymous referee for the many constructive suggestions that have
greatly improved the manuscript, Rachel Mandelbaum for carefully reading through
the manuscript, and Melanie Simet for helpful discussions. YZ acknowledges the
support by the National Key Basic Research and Development Program of China (No.
2018YFA0404504), National Science Foundation of China (11621303, 11873038,
11890692), the science research grants from the China Manned Space Project (No.
CMS-CSST-2021-A01, CMS-CSST-2021-B01), the National One-Thousand the National
One-Thousand Youth Talent Program of China, and the SJTU start-up fund (No.
WF220407220). WC acknowledges the support from the European Research Council
under grant 670193.  YZ thanks the inspiring discussions with Cathy Huang during
his visits to the Zhangjiang Hi-Tech Park.






\bsp	
\label{lastpage}

\end{document}